\documentclass[superscriptaddress,nofootinbib,longbibliography,onecolumn,notitlepage,aps]
{revtex4-2}
\pdfoutput=1
\usepackage{amsmath,amsfonts,amssymb,amsthm,amsbsy,mathtools}
\usepackage{graphicx}
\usepackage{dcolumn}
\usepackage{bm}
\usepackage{bbold}
\usepackage{mathtools}
\usepackage{xcolor}
\usepackage{comment}
\usepackage{booktabs} 
\usepackage{multirow}
\usepackage{graphicx}   
\usepackage{makecell}   
\usepackage{caption}
\usepackage{subcaption}
\usepackage{soul}
\usepackage{csquotes}
\usepackage{appendix}
\usepackage[normalem]{ulem}
\usepackage{xspace}
\usepackage{ifthen}
\usepackage[colorlinks,citecolor=blue,linkcolor=blue,urlcolor=blue, breaklinks=true]{hyperref}

\def\phys{{\mathrm{phys}}}
\def\kin{{\mathrm{kin}}}

\def\H{{\mathcal{H}}}
\def\CC{{\mathbb{C}}}
\def\ZZ{{\mathbb{Z}}}
\def\longto{{\longrightarrow}}

\newcommand{\ket}[1]{{|#1\rangle}}
\newcommand{\bra}[1]{{\langle#1|}}
\newcommand{\kket}[1]{{|#1\rangle\!\rangle}}
\newcommand{\bbra}[1]{{\langle\!\langle#1|}}
\newcommand{\matrixel}[3]{{\langle #1 | #2 | #3\rangle}}
\newcommand{\ketbra}[2]{{| #1 \rangle\!\langle #2|}}

\def\up{{\mathord{\uparrow}}}
\def\down{{\mathord{\downarrow}}}

\makeatletter
\let\@afterindenttrue\@afterindentfalse
\makeatother

\usepackage{xcolor}
\usepackage{xspace}
\usepackage{ifthen}
\newcommand{\showcomments}{true}

\newcommand{\acd}[1]%
{\ifthenelse{\equal{\showcomments}{true}}{{\color{magenta}{#1}}}{\xspace}}%

\definecolor{navy}{RGB}{0,70,180}

\newcommand{\andrea}[1]%
{\ifthenelse{\equal{\showcomments}{true}}{{\color{navy}{[#1]}}}{\xspace}}%

\usepackage{ifthen}
\usepackage[normalem]{ulem}
\newcommand{\trackchanges}{true} 

\definecolor{darkgreen}{RGB}{0,128,42}
\definecolor{darkred}{RGB}{179,0,0}

\newcommand{\removed}[1]{%
\ifthenelse{\equal{\trackchanges}{true}}%
{{\color{gray}{\small{\sout{#1}}}}}%
{\xspace}%
}
\newcommand{\added}[1]{%
\ifthenelse{\equal{\trackchanges}{true}}%
{{\color{teal}{#1}}}%
{#1}%
}%

\begin{document}

\author{Andrea Di Biagio}
\thanks{These authors contributed equally to this work.}
\affiliation{Institute for Quantum Optics and Quantum Information (IQOQI),
Austrian Academy of Sciences, Boltzmanngasse 3, A-1090 Vienna, Austria}
\affiliation{Basic Research Community for Physics e.V., Mariannenstraße 89, Leipzig, Germany}

\author{Anne-Catherine de la Hamette}
\thanks{These authors contributed equally to this work.}
\affiliation{Institute for Theoretical Studies, ETH Zürich, 8006 Zürich, Switzerland}
 \affiliation{Basic Research Community for Physics e.V., Mariannenstraße 89, Leipzig, Germany}

\title{Adlam's Frame: comment on ``Wigner's Frame''}

\date{July 20, 2026}

\begin{abstract}
   \noindent Recent no-go theorems based on extended Wigner's Friend scenarios (EWFS) reveal a deep tension between the universal validity of quantum theory and local friendliness (LF), the conjunction of three natural assumptions: the absoluteness of observed events, locality, and no-superdeterminism. In recent work~\cite{adlam2025wigners}, Adlam argues that this tension can be dissolved by appealing to quantum reference frames (QRFs).  We present and critically assess her proposal. We show that her proposed resolution does not arise from QRFs, but instead relies on three independent modifications: certain degrees of freedom are always definite, observers can flip upon measuring spins, and the outcomes the superobservers record and use to test the inequalities are not their friends' observed results. The proposal and the arguments for the plausibility of these modifications rest on several misconceptions about QRFs and EWFS, which we rectify.  We conclude that, within standard quantum theory, quantum reference frames do not evade the EWFS no-go theorems.   
\end{abstract}

\maketitle

\section{Introduction}

Thought experiments involving the unrestricted application of quantum theory to systems of arbitrary size have long been recognised as posing significant challenges to our physical intuitions~\cite{schroedinger1935gegenwaertige}. A recent wave of no-go theorems built on Wigner's original friend scenario~\cite{wigner1962remarks} has sharpened these challenges considerably. Beginning with Brukner's work~\cite{brukner2015quantum,brukner2018nogo} and culminating in Bong \textit{et al.}'s \cite{bong2020strong} no-go theorem, it has been shown that universal validity of quantum theory is incompatible with the conjunction of three natural assumptions: the absoluteness of observed events (AOE), locality, and no-superdeterminism~\cite{bong2020strong,cavalcanti2021implications}. Their conjunction is known as local friendliness (LF).

In the extended Wigner's Friend scenario (EWFS) considered by Bong \textit{et al.}~\cite{bong2020strong}, two friends\footnote{We use Adlam's naming convention rather than that of \cite{bong2020strong} to avoid confusion within this comment.} Alice and Chidi, each receive one half of an entangled bipartite quantum system, say, for concreteness, a pair of spin-1/2 particles. They each perform measurements on their spin inside isolated laboratories, recording outcomes $a$ and $c$. On each round of the experiment, the superobservers Bob and Divya perform one of several possible experiments on their respective friends and spins. The experiment is arranged such that Alice and Bob are spacelike separated from Chidi and Divya. Denote by $b$ and $d$ the measurement outcomes of Bob and Divya, and by $x$ and $y$ their measurement choice, respectively. Let $p(bd|xy)$ denote the statistics observed by Bob and Divya. 

The LF no-go theorem by Bong \textit{et al.}~\cite{bong2020strong} reveals a tension between the statistics of the superobservers' measurements and natural assumptions about observers' experiences and measurement choices. Indeed, it is natural to suppose that Alice and Chidi obtain a definite result on each round and that Bob and Divya simply \textit{discover} them when they ask them what they saw. This assumption is known as \textit{absoluteness of observed events} (AOE). It implies that $p(bd|xy)$ is the marginal over $a$ and $c$ of a joint probability distribution $P(abcd|xy)$ and that $b=a$ and $d=c$ in the rounds when Bob asks Alice what she saw and Divya asks Chidi what she saw, respectively. Bong \textit{et al.} have shown that the predictions of quantum theory are incompatible with AOE together with the assumption that Bob and Divya can make their measurement choices independently of one another and of their friends' measurement results. If, in particular, Alice and Bob are spacelike separated from Chidi and Divya during their measurements, the predictions of quantum theory are in contradiction with AOE, \textit{locality,} and \textit{no-superdeterminism}. 

This result has drawn a considerable amount of attention, as any way out of the contradiction appears to carry a significant cost. If one assumes that quantum theory can be successfully applied to arbitrary systems, then the no-go theorem either implies quantum theory fails at a certain scale, or that AOE fails, or that there are non-local, superdeterministic, or retrocausal correlations between observed events.

A number of authors take the LF no-go theorem, and Wigner's Friend thought experiments more broadly, as an indication that there are no single absolute, observer-independent events according to quantum theory \cite{wallace2014emergent,brukner2015quantum,rovelli1996relational,brukner2020facts,debrota2020respecting,cavalcanti2021implications,dibiagio2025relative,healey2025perspectives}. Emily Adlam, on the other hand, has been a vocal critic of relaxing AOE, on the grounds that it risks undermining scientific practice and thus the empirical support for quantum theory itself \cite{adlam2022does,adlam2023information,adlam2023doesnonabsolutenessobservedevents,adlam2024how,adlam2024moderate,adlam2025what}. Certainly, giving up AOE is a radical shift from classical ontology. However, it can be argued that communication and the practice of science need no revision: without AOE, intersubjective truth is still established by interacting with one another and with the world, much as objects can be at rest relative to one another even though there is no absolute notion of motion \cite{brukner2003information,zurek2003decoherence,zeilinger2005message,rovelli2024alice,covoni2026tractatus,dibiagio2025relative}. Even so, letting go of AOE remains a big shift in our intuitions about the world, and a proposal that preserves AOE without incurring any of the above-mentioned costs deserves close attention.

It is against this backdrop that Adlam writes \emph{Wigner's Frame}~\cite{adlam2025wigners}, in which she proposes a modification of quantum theory, inspired by the quantum reference frames programme, that preserves AOE but that, nevertheless, is supposed to reproduce all predictions of quantum theory, including the LF-violating correlations. In this proposal, the friend's measurement outcome is absolute and definite, while it is the \textit{orientation} of her laboratory relative to the superobserver's that becomes indefinite. This is allowed, she argues, because the kind of isolation required to perform an extended Wigner's Friend scenario prevents the agents from sharing a common reference frame. Adlam then claims that this fact dissolves the apparent tension between the universal applicability of quantum theory and AOE, defusing the no-go theorem: quantum theory does not apply to observations of quantities defined relative to frames in indefinite orientation relative to one another. More broadly, Adlam goes beyond the Wigner's Friend scenario and offers a general principle to decide which variables are described by quantum theory and which are not.

This appeal to indefinite frame relations is striking, because the formalism of quantum reference frames is by now well-developed within standard quantum mechanics and offers concrete tools precisely for situations in which agents do not share a common classical reference frame. The information-theoretic treatment of Bartlett, Rudolph and Spekkens~\cite{Bartlett_2007} shows how observers without access to a shared frame can nonetheless make consistent quantum mechanical predictions by appropriately averaging over the relevant symmetry transformations. More recent developments treat reference frames themselves as quantum systems, so that a frame may be in a coherent superposition with respect to another and frame changes between such frames can be defined and implemented within standard quantum theory; see, e.g.,~\cite{Giacomini_2017_covariance,Vanrietvelde_2018a,Hoehn_2019_trinity,delaHamette_2020, Castro_Ruiz_2021, Carette_2023}. Within these frameworks, well-defined probability distributions for relational observables are obtained even when the relation between frames is undefined.

In this comment, we carefully assess Adlam's proposal. Contrary to what one might assume on a first read, we find that the apparent resolution of the LF no-go result does not in fact arise from the use of the quantum reference frames formalism. Instead, it relies on three independent \emph{modifications}. Two are modifications of quantum mechanics, namely: (\textbf{A}) the degrees of freedom associated with observers' experiences are always definite; and (\textbf{B}) when a sufficiently isolated observer measures a symmetry-variant quantity, their own corresponding degree of freedom becomes entangled with it. The third is a modification of the experimental protocol: (\textbf{C}) the outcomes recorded by the superobservers in the LF statistics are not their friends' observed results, but the orientations of the spins relative to their own laboratories.

In Section~\ref{sec:Adlam}, we systematically present Adlam's modification of quantum theory and the ensuing treatment of (extended) Wigner's scenarios. In particular, we will see that Adlam's modified theory \emph{does} predict statistics exceeding the LF bounds but, we will argue, these do not constitute a genuine violation of local friendliness, as the experiments required to generate these statistics do not actually involve the friends' memory states (not even in the rounds where the superobservers ask their friends what they saw). In Section~\ref{sec:UQT}, we contrast Adlam's proposal with standard quantum theory. We first comment on Adlam's framing of the LF no-go theorem and clarify that the theorem constrains the statistics of the superobserver measurements, not joint predictions for the superobservers and the friends. We then correct the claim that quantum theory is unable to generate predictions for quantities defined relative to frames in coherent superpositions relative to each other. This can be done by using the tools developed in the QRF literature. Finally, we construct an explicit measurement model demonstrating that it is possible for the superobservers and the friends to keep their frames aligned  while performing the EWFS experiment, contrary to Adlam's claim. In Section~\ref{sec:Adash}, we briefly examine Adlam's proposal for a general principle relating symmetry and quantum definiteness and find it to encounter serious difficulties. We summarise our analysis and discuss its broader significance in Section~\ref{sec:discussion}.

\section{Adlam's proposal}
\label{sec:Adlam}

Adlam frames her proposal as a resolution of the LF no-go theorem based on quantum reference frames and symmetry considerations. She is open that it rests on additional postulates beyond standard quantum mechanics, but takes these to be natural extensions of the QRF research programme. We argue otherwise: Adlam's resolution relies on three independent modifications, of quantum theory and EWFS, and the apparent LF violations Adlam derives are predictions of this modified theory for a different experiment---\textit{not} of standard quantum mechanics together with the QRF formalism applied to EWFS. In what follows we collect these modifications, introduced at different points in her paper, into a single modified theory we may call Adlam's modified quantum mechanics (AMQM), and trace their consequences for a single Wigner's Friend scenario and for the EWFS. 

\subsection{Adlam's modifications of quantum theory}
\label{sec:AMQM}

Let us highlight and comment on the three modifications Adlam introduces  in her work~\cite{adlam2025wigners,adlam_email}.

\begin{itemize}\setlength\itemsep{0em}
	\item \textbf{A. Observers' experimental outcomes are always definite:} In AMQM, the degrees of freedom associated with an observer's experience or knowledge never go in superposition, thus they do not obey standard quantum theory.

	\item \textbf{B. Observers flip:} When sufficient isolation is present, and when an agent measures a symmetry-variant degree of freedom, \textit{their own} corresponding symmetry-variant degree of freedom may become entangled with it. Concretely, when measuring the orientation of an electron spin, the orientation of a sufficiently isolated laboratory becomes maximally entangled with that of the spin.
    
	\item \textbf{C. Substituted outcomes:} In the rounds of an EWFS in which the superobservers are supposed to learn their friends' outcomes, the outcome they record and use in the statistics is instead the orientation of the spin relative to \emph{their own} laboratory, which is statistically independent of the friend's outcome.
\end{itemize}
\noindent We remark that in fact Adlam proposes a stronger modification than \textbf{A}, namely,
\begin{itemize}\setlength\itemsep{0em}
	\item \textbf{A'. Symmetry-invariant degrees of freedom are always definite:} In AMQM, symmetry-invariant degrees of freedom of \enquote{well-integrated} systems never go in superposition, thus they do not obey quantum theory.
\end{itemize}
\noindent Since observers' experiences are related to internal configurations of their brains (which are the same whether translated, rotated, boosted, etc.), modification \textbf{A'} implies modification \textbf{A}.  We chose to distinguish between \textbf{A} and \textbf{A'} for the sake of exposition, as the ramifications of \textbf{A'} are not relevant to Adlam's analysis of Wigner's friend type scenarios; we come back to \textbf{A'} in Section~\ref{sec:Adash}.

\medskip

Modification \textbf{A} is sufficient to preserve AOE: since the physical variables associated with an observer's experience never undergo quantum indefiniteness, one can take their observations as absolute facts. Wigner himself advanced the view that it made no sense to think of an observer as being in a superposition of having witnessed different outcomes and that the friend's observation should correspond to a definite outcome~\cite{wigner1962remarks}. Famously, Penrose proposed a direct link between consciousness and the collapse of the wavefunction~\cite{hameroff2014consciousness}, and developing such an alternative to quantum theory is an active research direction~\cite{chalmers2021consciousness}. Spontaneous collapse theories such as GRW \cite{ghirardi1986unified} and Di\'osi-Penrose \cite{diosi1989models} also implement modification \textbf{A}~\cite{bassi2013models}.\footnote{We note that the de Broglie-Bohm theory satisfies AOE but \textit{not} modification \textbf{A}, as observers' states of mind are still assigned superposed quantum states. }

If one's aim was simply to ensure AOE one could stop here. But note that with only modification \textbf{A} in place, the EWF experiment cannot be done: no violation of an LF inequality can be observed. This  is because the quantum protocol relies on the superobservers performing an interference experiment on the friends' memory. Yet, in Adlam's proposal there is an \textit{apparent} violation%
\footnote{Adlam herself never speaks of violations of the LF inequalities, only of reproducing \enquote{the standard quantum predictions for measurements in the orientation basis}~\cite[p.\,10]{adlam2025wigners}; we comment on the relation between these two ways of speaking in Section~\ref{sec:variables_in_EWFS}. Throughout this comment, we say that a theory produces an \textit{apparent} violation of the LF inequalities when it predicts superobserver statistics $p(bd|xy)$ exceeding the LF bounds---so that experimenters running the protocol would report a violation---while the friends' observed outcomes never enter the reported statistics.} %
of the local friendliness inequalities: physicists performing the (\textbf{C}-modified) EWFS experiment are predicted to observe stronger-than-classical correlations, even though the friends' memories in the experiments never go in a superposition of different observations. This is achieved via modifications \textbf{B} and \textbf{C}. 

\medskip

Modification \textbf{B} concerns the orientation of the friends' labs. When an observer measures a spin in a sufficiently isolated lab (and obtains a definite outcome according to \textbf{A}) the orientation degree of freedom \textit{of their entire lab} now enters a quantum superposition, entangled with the spin.
Adlam motivates adding \textbf{B} as doing \enquote{justice to the intuition} that measurement outcomes are in some sense indefinite, without resorting to \enquote{relative facts or sacrificing the reliability of communication, memory or language} \cite[p.\,24]{adlam2025wigners}. 
She claims that to perform an EWF experiment, the friends \textit{have} to be isolated to such a degree that they would not be able to tell that they are now in a new or indefinite orientation relative to an external observer.

Note that modifications \textbf{A} \textit{and} \textbf{B} are not sufficient for the \textit{apparent} violation of the LF inequalities. This is where modification \textbf{C} comes in.

\medskip

According to modification \textbf{C}, on the rounds in which a superobserver ``opens the box'' of the isolated laboratory to ask their friend what they saw, the superobserver \emph{does not} record the friend's outcome. Instead, they record the orientation of the spin relative to their lab (perhaps by directly performing a quantum measurement on the spin's orientation, or by combining a measurement of the \textit{friend}'s orientation and the friend's recorded outcome) and use this to compute the statistics for the EWFS. It is only because of this modification of the protocol and the indefiniteness in the \textit{orientation} degrees of freedom of the friend that it is possible for the superobservers to report \textit{apparent} local friendliness inequality violations.

This is a modification because the LF no-go theorem requires the superobservers to ask their friend for the outcome of their measurement and use that in the computation of the correlations. Indeed, this is the \textit{only} constraint on the kind of measurement performed by the superobservers on their friends. We will review the LF no-go theorem in Section~\ref{sec:variables_in_EWFS}, where we also comment on the reasons why Adlam introduces modification \textbf{C}.

\subsection{Adlam's Friend}
\label{sec:AF}

Let us follow Adlam's treatment of a Wigner's Friend scenario to see how these modifications work in practice. The friend, Alice,  is an observer in an isolated lab who is about to perform a spin measurement. Her frame of reference is initially aligned with that of Bob, who will play the role of the superobserver (Wigner). Adlam distinguishes between three variables pertaining to Alice, namely, $A_I$, $A_R$, and $A_E$. 

When Alice measures the spin's orientation, she can find it either aligned or anti-aligned to an axis in her lab and this outcome is represented by the variable $A_I$ (where $I$ stands for \textit{internal}). According to modification \textbf{A}, this variable always has a definite value. The variable $A_R$ is the \textit{orientation} of Alice's lab with respect to Bob's lab. This is the variable that, according to modification \textbf{B}, becomes maximally entangled with the spin during Alice's measurement. Finally, $A_E$ is defined as Alice's result, but translated into Bob's frame, with $E$ standing for \textit{external}. When $A_I$ and $A_R$ are both definite, then $A_E = A_I\times A_R$.

According to the standard quantum mechanical treatment, Alice's measurement would be modelled by Bob as a unitary operation entangling the spin and $A_I$,
\begin{equation}\label{WF_standard}
  \frac1{\sqrt2} \left(\ket\up+\ket\down\right)_S\otimes\ket{0}_{A_I}\ket{\up}_{A_R}
  \;\longto\;
   \frac1{\sqrt2}\left(\ket\up_S\ket{{+}1}_{A_I}+\ket\down_S\ket{-1}_{A_I}\right)\otimes\ket{\up}_{A_R},
\end{equation}
where $A_I=0$ denotes that Alice has not observed the result of the spin measurement yet and $A_I={+}1$ or $-1$ correspond to having observed the spin being aligned or anti-aligned to her axis, respectively. Alice's frame remains aligned with Bob's during the measurement.

In AMQM, however, $A_I$ remains definite at all times and Alice \textit{either} sees aligned \textit{or} anti-aligned. It is instead  Alice's \textit{orientation} that becomes entangled with the spin.
So the evolution is \textit{either}
\begin{equation}\label{evo_P_our_notation}
  \frac1{\sqrt2} \left(\ket\up+\ket\down\right)_S\otimes\ket{\up}_{A_R}\ket{0}_{A_I}
  \;\longto\;
   \frac1{\sqrt2}\left(\ket\up_S\ket{\up}_{A_R}+\ket\down_S\ket{\down}_{A_R}\right)\otimes\ket{{+}1}_{A_I},
\end{equation}
in the case where Alice observes the spin to be aligned to her choice of axis, \textit{or}
\begin{equation}\label{evo_A_our_notation}
  \frac1{\sqrt2} \left(\ket\up+\ket\down\right)_S\otimes\ket{\up}_{A_R}\ket{0}_{A_I}
  \;\longto\;
   \frac1{\sqrt2}\left(\ket\up_S\ket{\down}_{A_R}+\ket\down_S\ket{\up}_{A_R}\right)\otimes\ket{-1}_{A_I},
\end{equation}
in case she finds the spin to be anti-aligned. Two comments are in order.

\medskip

The first comment is about notation. The notation we are using is introduced by Adlam in footnote 3 of~\cite{adlam2025wigners}. In the paper she mostly uses a notation that does not explicitly assign a state to $A_I$ and $A_R$ separately, and instead writes the final state of evolution~\eqref{evo_P_our_notation} as
\begin{equation}\label{P_Adlam}
  \frac1{\sqrt2}\ket\up_S\ket{P}_A+\frac1{\sqrt2}\ket\down_S\ket{P'}_A,
\end{equation}
where ``$\ket{P}_A$ represents a state of Alice where she has observed the particle
to be parallel to her own axis; $\ket{P'}_A$ also represents a state of Alice where she
has observed the particle to be parallel to her own axis, but it has a different
total orientation relative to Bob''~\cite[p.~8]{adlam2025wigners}. Adlam prefers this notation because ``one possible interpretation of [her proposal] is that `what Alice has seen' should not be treated as a quantum degree of freedom at all, in which case it should not be represented as a quantum state in a Hilbert space''~\cite[p.~9]{adlam2025wigners}.

We prefer our notation for two reasons. The first reason is that assigning a Hilbert space to a system does not automatically make it a quantum system. Several approaches to classical-quantum hybrid dynamics indeed do associate Hilbert spaces to classical systems, with classicality enforced by a restriction to a commutative subalgebra of observables, or to states diagonal in a privileged basis~\cite{oppenheim2022two, oppenheim2023postquantum}. The second reason is clarity. Alice's lab does not have a \textit{merely} different total orientation relative to Bob in $\ket{P}_A$ compared to $\ket{P'}_A$: it has the \textit{opposite} orientation. Indeed, since $\ket\down_S$ means that the spin is anti-aligned to Bob's frame and since ``in the process of performing the measurement Alice aligns her own laboratory with $S$''~\cite[p.\,16]{adlam2025wigners},  the state $\ket\down_S\ket{P'}_A$ means that Alice's frame is anti-aligned to Bob's frame.\footnote{Note this also follows from the definition, $A_E=A_I\times A_R$: $A_E$ can only take the values $\pm1$, so $A_R$ can only take values $\pm1$, that is, up or down.} Writing the state as~\eqref{P_Adlam} obfuscates this, while also making it harder to read which variables are entangled and which ones factor out.

\medskip

The second comment is about dynamics. Adlam does not specify which of the two evolutions happens, only that \textit{either} one \textit{or} the other happens on each round, ``perhaps selected at random''~\cite[p.~8]{adlam2025wigners}.
Additionally, it is not specified whether or how the choice of evolution might be affected by the quantum state of the spin. For concreteness, and since all quantum states she writes down are equal superpositions of eigenstates of the measured observable, we will proceed by assuming that one of the two evolutions happens at random with equal probability.

\medskip

We now come to Bob's measurement. In the standard treatment, when Bob wants to know what Alice saw, this is represented by a quantum measurement of the variable $A_I$ on the final state in~\eqref{WF_standard}, which yields ${+}1$ or $-1$ with probability $1/2$. According to QM, the orientation $A_R$ of Alice’s laboratory need not change during her measurement of the spin (see Section~\ref{sec:isolation_shared_reference}) and Bob may verify this. In AMQM, the situation is different. Bob can ask Alice what she saw, because of modification \textbf{A} her outcome $A_I$ is definite on every round, and asking simply reveals it.  He may also measure the orientation $A_R$ of Alice's lab, and find that it has flipped on half of the rounds, as per modification \textbf{B}. Finally, he may measure $A_E$, the spin orientation in \textit{his} frame,  and find it to be up or down with 50\% probability. Note that in both  final states of~\eqref{evo_P_our_notation} and~\eqref{evo_A_our_notation}, $A_E$ and $A_R$ are entangled in such a way as to satisfy $A_I\times A_R=A_E.$

Modification \textbf{C} concerns only the protocol in the extended Wigner's Friend scenario, as we discuss next.

\subsection{Extended Adlam's Friend scenario}
\label{sec:EAFS}

\begin{table}[h]
{\centering
\begin{tabular}{@{}ccccc@{}}
\toprule
Work & Theory & Variable entangled with spin & Variable measured when $x,y=1$ & Variable measured when $x,y\neq1$ \\
\midrule
Bong \textit{et al.}~\cite{bong2020strong}        & QT  & friend's memory      & friend's memory & spin and friend's memory \\
Adlam~\cite{adlam2025wigners}       & AMQM & friend's orientation & spin orientation & spin and friend's orientation \\
\bottomrule
\end{tabular}\par}
\captionsetup{justification=justified}
\caption{\leftskip=0pt\rightskip=0pt  Comparison of two thought experiments. The LF-inequality-violating protocol proposed by Bong \textit{et al.}~in~\cite{bong2020strong} is analysed assuming the universal validity of quantum theory. In their protocol, the friends' memory degrees of freedom become entangled with the spins. When $x=1$ or $y=1$, the corresponding superobserver probes the memory of their friend, asking them what they saw, otherwise they perform an interference experiment on their friend and the spin. Adlam, in~\cite{adlam2025wigners}, considers a similar-looking protocol within her modified theory (AMQM). Per modifications \textbf{A} and \textbf{B}, it is the friends' \emph{orientation} degrees of freedom that become entangled with the spins, and, per modification \textbf{C}, the $x=1$ and $y=1$ measurements probe the orientation of the spin in the superobserver's own lab rather than the friend's memory. When the superobservers perform the interference experiment, they do so on the friends' orientation degrees of freedom, not on the memories. The apparent LF violation in Adlam's analysis is therefore a prediction in a different theory about a different experiment. On her own state assignment the memory registers always factorise, so no genuine LF violation can occur.}
\label{tab:bong-vs-adlam}
\end{table}

Let us now consider an extended Wigner's Friend scenario involving four agents, Alice, Bob, Chidi, and Divya, where Alice and Chidi play the role of the friends and Bob and Divya the role of the superobservers. At the beginning of the experiment, all four agents’ reference frames are aligned. A pair of spin-1/2 systems\footnote{Adlam calls them $X$ and $Y$. We refer to them as $S_1$ and $S_2$ since we will reserve $x$ and $y$ for the superobservers' measurement choices.} $S_1$ and $S_2$ is prepared in the maximally entangled state
\begin{equation}
 \ket{\Phi}_{S_1S_2} = \frac1{\sqrt2}\ket{\up\up}+\frac1{\sqrt2}\ket{\down\down}
\end{equation}
with respect to the initial orientation of the four agents' frame. Alice and Chidi now perform a measurement of $S_1$ and $S_2$ in their frame, respectively.

In the standard treatment, the friends' frames remain aligned with the superobservers' while their internal degrees of freedom become entangled with the spins, resulting in the state
\begin{equation}\label{EWFS_standard}
   \frac1{\sqrt2} \left(\ket{\up\up}_{S_1S_2}\ket{{+}1}_{A_I}\ket{{+}1}_{C_I} 
    + \ket{\down\down}_{S_1S_2}\ket{-1}_{A_I}\ket{-1}_{C_I}\right)\otimes\ket{\up}_{A_R}\ket{\up}_{C_R}.
\end{equation}
Afterwards, the superobservers perform their measurements. Let us call Bob's measurement choice $x$ and Divya's $y$. If $x=1$, Bob  ``opens the box'' and asks Alice what she saw; if $x\neq1$, then Bob performs an interference measurement on Alice and $S_1$. Call $b$ his outcome in either case. Divya does the same depending on the value of $y$, obtaining the outcome $d$. For appropriate choices of measurements, the statistics $p(bd|xy)$ predicted by quantum theory violate the local friendliness inequalities~\cite{bong2020strong}.

\medskip

Due to modifications \textbf{A}, \textbf{B}, and \textbf{C}, the experiment unfolds quite differently in AMQM. Because of modification \textbf{A}, the degrees of freedom associated with Alice's and Chidi's experience do not go in a superposition and, on each round, each friend definitely observes their spin to be aligned or anti-aligned to an axis in their lab. Meanwhile, because of modification \textbf{B}, Alice's and Chidi's labs' \textit{orientations} go into an entangled quantum superposition, depending on what they observe. Adlam writes only one of the four possible states in her notation as
\begin{equation}\label{PA+PpAp}
    \frac1{\sqrt2} \left( \ket{\up\up}_{S_1S_2}\ket{P}_A\ket{A}_C +\ket{\down\down}_{S_1S_2}\ket{P'}_A\ket{A'}_C \right).
\end{equation}
For completeness, let us write in our notation the four possible states after the friends' measurements,
\begin{equation}\label{eafs-states-new}
\begin{aligned}
\!\!\ket{\Psi_{++}} &= \frac{1}{\sqrt{2}} \bigl(\ket{\up\up\up\up}+\ket{\down\down\down\down}\bigr)_{S_1A_RS_2C_R}\ket{{+}1}_{A_I}\ket{{+}1}_{C_I}\,,~~~~
\ket{\Psi_{+-}} = \frac{1}{\sqrt{2}} \bigl(\ket{\up\up\up\down}+\ket{\down\down\down\up}\bigr)_{S_1A_RS_2C_R}\ket{{+}1}_{A_I}\ket{{-}1}_{C_I}\,,\\
\!\!\ket{\Psi_{-+}} &= \frac{1}{\sqrt{2}} \bigl(\ket{\up\down\up\up}+\ket{\down\up\down\down}\bigr)_{S_1A_RS_2C_R}\ket{{-}1}_{A_I}\ket{{+}1}_{C_I}\,,~~~~
\ket{\Psi_{--}} = \frac{1}{\sqrt{2}} \bigl(\ket{\up\down\up\down}+\ket{\down\up\down\up}\bigr)_{S_1A_RS_2C_R}\ket{{-}1}_{A_I}\ket{{-}1}_{C_I}\,.
\end{aligned}
\end{equation}
On each round, therefore, after the friends have performed their measurements, the state of the spins, the friends, and their labs relative to the superobservers is one of these four states. 

Note that, given a statistical mixture of these four states, it is impossible to violate the LF inequalities by performing measurements via the Bong \textit{et al.} protocol described above. This is because there is no entanglement between the friends' memories $A_I$ and $C_I$, as they are in a definite state in each round.

Instead, in Adlam's version of the experiment, on the rounds in which the superobservers are supposed to learn about their friend's observed outcomes, they record the spin orientation in their own frame. That is, on the rounds where $x=1,$ Bob measures $A_E$ instead of $A_I$ (he may do this either by measuring both $A_I$ and $A_R$ and computing $A_E = A_I \times A_R$, or by directly measuring the spin $S_1$), and similarly for Divya when $y=1$. This is modification \textbf{C}.

Although she does not cover this explicitly, in the other rounds, the superobservers must perform an interference experiment on the spins and the friends' \textit{orientations}, rather than the spins and the friends' \textit{memories} as in the standard treatment.

\medskip

At this point Adlam claims that each of the states in \eqref{eafs-states-new}
\begin{quotation}
	\noindent\textit{\enquote{is one in which the observed outcomes are absolute and yet it also reproduces the standard quantum predictions for measurements in the orientation basis, without retrocausality or superdeterminism. So this approach seems to offer an appealing way of solving the problem posed by the EWF scenarios without sacrificing anything essential in the process}}~\cite[p.\,10]{adlam2025wigners}. 
\end{quotation}
Given that this is the keystone of the proposal, it is worth analysing in detail. 

First, these states plainly do not reproduce ``the standard quantum predictions for measurements in the orientation basis''. This is only true for measurements on the \textit{spins} $S_1$ and $S_2$. However, according to quantum theory both $A_R$ and $C_R$ remain aligned with the external frame during the friends' measurement while, according to \eqref{eafs-states-new}, there will be a significant chance of finding one of the friends' orientation to be inverted!

Second, the EWF paradox is not about the quantum predictions on orientation degrees of freedom, but on the memory degrees of freedom, which in these states are not in a superposition.
So one needs to ask what Adlam actually means when she says that ``this approach seems to offer an appealing way of solving the problem posed by the EWF scenarios''. What Adlam means, as already anticipated in Section~\ref{sec:AMQM}, is that \emph{thanks to modification \textbf{C}}, the superobservers can obtain correlations stronger than the LF bounds while preserving AOE. However, this is just the \textit{appearance} of LF violations, since the EWFS experiment is supposed to be performed on the friends' memories, while in her version of the experiment, via \textbf{C}, the outcomes that go in the computation of the inequalities never depend on the friends' internal states.

Finally, we must note that modification \textbf{C} \emph{does} sacrifice something essential: the connection between the experiment performed and the experiment the no-go theorem is about.

\medskip

Incidentally, note that since the superobservers obtain one of the four different possible states in each round, they are effectively dealing with a \textit{statistical mixture} in which there is no entanglement between $A_I$, $C_I$, and the rest of the systems. And since Adlam's discussion remains at the level of state assignments and variables, one may wonder if the superobservers can actually violate the LF inequalities. Indeed, if the superobservers were performing the measurements in the Bong \textit{et al.}~protocol, they would not be able to violate any inequality. Surprisingly enough, thanks to modification \textbf{C}, one can obtain an \textit{apparent} violation and, in fact, saturate Tsirelson's bound. We show this in Appendix~\ref{app:violation}.  In short, the only way to achieve the apparent LF inequality violations in AMQM is for Bob and Divya to \textit{not} just probe $A_I$ and $C_I$, and instead perform measurements on $S_1A_R$ and $S_2C_R$.

Note that reaching the Tsirelson bound indeed does not constitute a violation of \textit{local friendliness}: the local friendliness assumptions, applied to Adlam's version of the experiment---an experiment that only involves measurement results on orientation degrees of freedom---yield a different bound than they yield on the actual extended Wigner's friend experiment. In fact, the LF assumptions applied to Adlam's protocol merely imply no-signalling correlations \cite[Methods]{bong2020strong}.

\medskip

In summary, modification \textbf{A} makes sure that the friends' experiences are definite at all times. This modification automatically makes it impossible to violate the local friendliness inequalities in an EWF scenario. Adlam's modifications \textbf{B} and \textbf{C} add the possibility of the \textit{appearance} of LF violations at the cost of introducing a stochastic \emph{flipping} of isolated observers, as well as the superobservers performing an experiment involving different degrees of freedom altogether.

\section{How quantum theory makes predictions for variables in different frames}
\label{sec:UQT}

In this section, we examine several claims Adlam makes about quantum theory and reference frames.

We first consider the variables featuring in the no-go theorem, and clarify that the Bong \textit{et al.}~\cite{bong2020strong} result is a constraint on the observed statistics $p(bd|xy)$, regardless of how these are interpreted in terms of internal or external variables. Second, we address Adlam's claim that quantum theory does not offer predictions for variables defined in different frames. Third, we discuss the requirement of isolation between friends and superobservers in EWFS and provide an explicit von Neumann measurement model demonstrating that a shared reference frame can be maintained even with the sort of isolation required in an EWFS.

\subsection{The relevant variables in Bong \textit{et al.}}
\label{sec:variables_in_EWFS}

A reader coming from Adlam's paper may be surprised that we speak of local friendliness violations in this analysis, since the term appears nowhere in Adlam's work. Indeed, Adlam does not talk about the LF no-go theorem as imposing bounds on the correlations between the two superobservers Bob and Divya, and instead presents the theorem as concerning the quantum mechanical predictions for four pairs of (super)observers (Alice and Chidi, Alice and Divya, Bob and Chidi, and Bob and Divya). Here we relate these two formulations of the theorem and remark on how Adlam uses her framing to introduce modification \textbf{C}.

\medskip

Let us briefly summarise the standard presentation of the LF no-go theorem~\cite{bong2020strong}. The no-go theorem concerns bounds on the correlations $p(bd|xy)$ of the superobservers' outcomes imposed by some natural assumptions. As we mentioned, \textit{absoluteness of observed events} (AOE) implies that the data collected by the superobservers can be obtained as a marginal 
\begin{equation}\label{aoe}
  p(bd|xy) = \sum_{ac}P(abcd|xy)
\end{equation}
over a joint probability distribution of all the observers' results and that the superobservers can learn about the results of their friends, namely, that the joint distribution $P$ satisfies
\begin{equation}\label{tracking}
  P(b|acdy,x{=}1)=\delta_{ab}, ~~~~~P(d|abcx,y{=}1)=\delta_{cd}.
\end{equation}
The assumption of \textit{locality} implies that each superobserver's outcome is independent of the distant superobserver's setting, that is,
\begin{equation} \label{loc}
  {P(b|acxy) = P(b|acx)},~~~~~~~~~{P(d|acxy) = P(d|acy)}.
\end{equation}
 Finally, assuming \textit{no-superdeterminism} implies that the friends' outcomes are uncorrelated with the superobservers' measurement settings, that is,
\begin{equation} \label{nsd}
   P(ac|xy)=P(ac).
\end{equation}
 Together, AOE, locality, and no-superdeterminism imply bounds on the observed correlations $p(bd|xy)$, expressed in terms of so-called \textit{local friendliness} (LF) inequalities.\footnote{Note that the single assumption of \textit{local agency}, which states that \enquote{[t]he only relevant events correlated with an intervention are in its future light cone}, is in fact sufficient to derive \eqref{loc} and \eqref{nsd} \cite{wiseman2017causarum,cavalcanti2021implications}. Note also that some authors \cite{schmid2024review,riedel2024quantum} have argued that, strictly speaking, Eq.~\eqref{tracking} does not follow from AOE alone, but requires an additional, implicit, assumption called \textit{tracking}: \enquote{an observer can faithfully transmit their observations to other observers}~\cite[p.~15]{schmid2024review}.}

Note that the theorem places no constraint whatsoever on \emph{what} the friends measure. The friends perform some measurement (a spin component, the occupation number of a mode, whether an atom has decayed, or, for that matter, the orientation of a spin relative to some frame). The superobservers choose among a set of measurements, subject to a single constraint: one of the available choices consists of \emph{copying the friend's outcome}. No translation of that outcome into anybody's reference frame is involved: the outcome is complete as it stands, and copying it is a frame-independent operation. Reference frames, orientations, and symmetries appear nowhere in the theorem.

\medskip

Let us compare this to how Adlam frames the result starting\footnote{A possible source of confusion is that Adlam's exposition on page 3 actually contains a misrepresentation of the no-go theorem: the presentation there describes a scenario in which the superobservers never ask the friends what they saw; the version with the choice is introduced on page 4.} on p.~4. She says that on some runs Bob asks Alice for her result $A$ and on other runs he performs an interference experiment on Alice, obtaining a result $B$. Divya does the same to Chidi, obtaining either $C$ or $D$. Adlam then presents the Bong \textit{et al.} no-go theorem as saying that there is no joint distribution over the four variables $A,B,C,$ and $D$ that reproduces the predictions of quantum theory for the four pairs $(A,C),(A,D),(B,C),$ and $(B,D)$ and which also satisfies corresponding causality assumptions.

At first sight, this formulation appears quite different from the one above. One involves only the statistics of the superobservers' measurements, while the other involves the predictions over four different pairs of variables. However, in the dichotomic case (two measurement choices and two possible outcomes per superobserver), the two can be shown to be equivalent and amount to a change of notation, as follows.

Identify $A$ with $b$ when $x=1$ and $B$ with $b$ when $x=2$, and similarly identify $C$ and $D$ with $d$ when $y=1$ and $2$, respectively. Then, when Adlam speaks of the predictions of quantum theory for different pairs of variables, she is speaking about the conditional probability $p(bd|xy)$ for different values of $x$ and $y$. For example, the quantum mechanical predictions for the pair $(A,D)$ are the statistics for $p(bd|x{=}1,y{=}2)$. This is perhaps unusual, but not unprecedented.\footnote{Indeed Arthur Fine showed that for any no-go theorem about conditional probabilities $p(ab|xy)$ there is a no-go theorem about the existence of a joint distribution $P(A_1A_2\dots B_1B_2\dots)$ reproducing several marginals $P(A_iB_j)$, where the identification is of the form $P(A_iB_j)\longleftrightarrow p(ab|x{=}i,y{=}j)$~\cite{fine1982hidden, fine1982joint}.} Stated in this way, whenever the statistics $p(bd|xy)$ violate the LF inequalities, there is no joint probability distribution over the four variables $A, B, C, D$ satisfying analogous properties.

\medskip

This presentation of the theorem allows Adlam to frame her argument. Indeed, on page 6, Adlam makes two assertions. First, she says that there is an ambiguity about the definition of the variable $A$, and that it could be \textit{either} $A_I$ (the outcome of Alice's observation) \textit{or} $A_E$ (the spin orientation in Bob's frame), and so it is unclear whether the no-go theorem is about a joint distribution over the variables $(A_I,B,C_I,D)$ or over the variables $(A_E,B,C_E,D)$. Adlam continues by arguing that the no-go theorem \textit{ought} in fact to be understood as referring to the four variables $(A_E,B,C_E,D)$ since \enquote{[w]e cannot possibly compare the distribution over $A_I, B, C_I, D$ to the predictions of quantum mechanics for a pair such as $(A_I, D)$ [... b]ecause quantum mechanics makes predictions for correlations between these pairs of variables only where both variables are stated \emph{relative to the same reference frame}} \cite[p.\,7, emphasis in the original]{adlam2025wigners}.

Neither assertion holds, for the following reasons. Firstly, quantum theory has no problem making predictions for variables stated relative to two different reference frames. Indeed, many techniques have been developed in the study of quantum reference frames; we discuss this further in the next subsection.

But, more importantly, there is no such ambiguity in the statement of the no-go theorem. When $x=1$, Bob is supposed to enter the lab, ask Alice what she saw, and record her answer: his outcome is, by definition, a copy of $A_I$.  Even if the experiment performed by Alice involves a quantity defined relative to a frame, her \textit{outcome} is a \textit{frame-independent data point} that requires no translation in order to be copied.  

 The statistics collected as per Adlam's proposal are then not the statistics the theorem bounds, and the fact that they saturate the Tsirelson bound carries no implication for local friendliness: the experiment performed is not an extended Wigner's friend experiment in the sense of Bong \textit{et al.}

\subsection{`No fact of the matter' and quantum reference frames}
\label{sec:nofact}

A central premise in Adlam’s analysis of the modified EWFS is that there may be \enquote{no fact of the matter} about the orientation of one observer’s reference frame relative to another’s. She argues that when there is no fact of the matter, \enquote{quantum mechanics will have nothing to say about what correlations we should expect to see} \cite[p.\,7]{adlam2025wigners}. However, this is not the case, as the formalism of quantum reference frames provides exactly the predictions Adlam says are impossible, as we now explain.

First, let us note that there being no fact of the matter is exclusively an issue introduced by Adlam via modification \textbf{B}. In standard quantum theory, for all practical purposes, Alice's lab does in fact not go out of alignment with Bob's frame when she measures a spin, as will become apparent in Section~\ref{sec:isolation_shared_reference}. But suppose, for the sake of the argument, that modification \textbf{B} holds.
In that case, the frameworks of quantum reference frames  provide systematic tools for describing and computing probability distributions over variables defined in reference frames whose relative orientation is not classically defined \cite{Bartlett_2007,Giacomini_2017_covariance,Vanrietvelde_2018a,Hoehn_2019_trinity,delaHamette_2020, Castro_Ruiz_2021, Carette_2023}. 

First, if the relative orientation between two frames is known, then all variables defined in one frame can simply be transformed by the corresponding rotation to the variables in the other frame. Correlations between variables are then simply evaluated on states defined in either frame.

Second, suppose the relative orientation is described by a coherent superposition. This is the \enquote{no fact of the matter} case considered by Adlam.  While there is no definite \emph{classical} orientation, this is a pure quantum state and it therefore corresponds to maximal information in the quantum mechanical sense: the superobserver has maximal knowledge of the relation between their own frame and their friend's. The situation in which a frame is in coherent superposition of orientations relative to another frame is the focus of a great deal of recent QRF literature~\cite{Giacomini_2017_covariance, Vanrietvelde_2018a, Hoehn_2019_trinity, delaHamette_2020}, where QRF transformations are regularly used to relate the descriptions of quantum systems relative to different frames.

Let us go into a bit more detail. Let $V^{A\to B}$ denote the frame change map transforming the state from the description relative to frame $A$ to that of frame $B$ in the perspectival or perspective-neutral formalism. That is, if $\ket\psi_{|A}$ is the state in $A$'s frame, then $\smash{\ket\psi_{|B}=V^{A\to B}\ket\psi_{|A}}$ is the state in $B$'s frame. This map also allows us to relate observables expressed in different frames. Indeed, let $O_{1|A}$ be an observable defined in frame $A$, then
\begin{equation}
  O_{1|A}^{A\to B}=V^{A\to B}O_{1|A}V^{B\to A}
\end{equation}
is the corresponding observable in $B$'s frame.  We can then compute correlations $C_\psi(O_{1|A}O_{2|B})$ between an observable $O_{1|A}$ in $A$'s frame and a second one $O_{2|B}$ in $B$'s by computing, in either frame,
\begin{equation}
  C_{\psi}(O_{1|A}O_{2|B})=\bra\psi O_{1|A}O_{2|B}^{B\to A}\ket\psi_{|A}=\bra\psi O_{1|A}^{A\to B}O_{2|B}\ket\psi_{|B},
\end{equation}
where we assumed, for simplicity, that $\bra\psi O_{1|A}\ket\psi_{|A}= \bra{\psi} O_{2|B}\ket{\psi}_{|B}=0$. We refer the reader to Appendix~\ref{app:QRFs} for a more detailed treatment.

Finally, note that QRF tools allow one to make predictions even in the case where there simply is \textit{no information whatsoever} about the relative orientation of the two frames. The relation between the frames would be represented by a totally mixed state and the ``twirling'' procedure of \cite{Bartlett_2007} would still allow for predictions about variables defined in those two frames. 

\medskip

A key insight to keep in mind here is that ``a variable stated in a reference frame'' is a single symmetry-invariant observable. For intuition, consider classical particles on a line with translation invariance. The position $x_A$ of particle $A$ in particle $B$'s reference frame is nothing but the distance $x_A-x_B$ between $A$ and $B$, which is an invariant. If we want to know the correlations between $x_{A|B}$ and $x_{C|D}$, variables ``stated with respect to two different frames'', all we need to do is to look at the correlations between $x_A-x_B$ and $x_C-x_D$, two symmetry-invariant quantities.

\subsection{Isolation with shared reference frames}
\label{sec:isolation_shared_reference}

Another central premise in Adlam’s analysis is that the kind of isolation required in an EWFS precludes the agents from maintaining a shared reference frame throughout the experiment. According to this view, any attempt by the agents to verify or preserve the alignment of their frames would necessarily decohere the quantum system being measured. However, this premise is incorrect.

Indeed, in the standard formulation of EWFS, all agents are \textit{assumed} to share a common reference frame at all times. 
In fact, in the quantum protocol used to violate the LF inequalities, the friends measure the spins in a fixed basis and the superobservers' interference experiment effectively measures the spins in a complementary basis. This is impossible without sharing a reference frame. In practice, such a shared frame can be fixed and continuously monitored by reference to a shared external field, such as the gravitational field of the Earth, or some pre-aligned gyroscopes.

Moreover, a friend and their superobserver can continuously be in contact about the alignment of their frames. It is well-known in the literature that a friend in a WF or EWF scenario can in fact communicate a great deal of substantive information to their superobserver without compromising the protocol~\cite{Deutsch_1985,brukner2015quantum, brukner2018nogo,Ying2024relatingwigners,DelSanto2025_wigner}. For example, Deutsch noted that the friend can announce \emph{that} she has performed a measurement and obtained a definite outcome  without disturbing the superposition, as long as she does not reveal \emph{which} outcome~\cite{Deutsch_1985}. In fact, Friend could communicate which measurement they performed, or Wigner could tell Friend what measurement they will decide to do. More generally, as long as the data exchanged does not contain information about the result of the experiment, it can be shared without spoiling the superposition.

It is only in AMQM, because of modification \textbf{B}, that the orientation of Friend's frame becomes entangled with the spin and, as a consequence, sharing information about Friend's orientation would decohere the superposition of Friend's lab and spin.

\medskip

In the following, we show that in standard quantum theory, it is possible for the friend to measure the spin \emph{without} their frame becoming entangled with it. The orientation of the friend's lab thus constitutes information that \emph{can} be freely shared between Friend and their superobserver throughout the experiment.
 
We do so by constructing a von Neumann measurement model in which a memory register records whether a spin-1/2 system is found to be aligned or anti-aligned with respect to a physical reference frame. We will see that this induces \emph{some} entanglement between the reference frame and the memory register. Thus, after the measurement, there is \textit{some} uncertainty about the relative orientation between this frame and the external one, and \emph{some} decoherence of the memory when attempting to communicate the orientation of the reference frame to other agents. Crucially, however, this entanglement can be made arbitrarily small by increasing the size of the friend's reference frame. In the limit of an arbitrarily large frame, no decoherence occurs.

Consider three systems: a spin-1/2 system $S$, a spin-$j$ system $R$ which serves as a reference frame,\footnote{Using a large spin-$j$ as an orientation quantum reference frame has been considered for example in~\cite{poulin2006toy} and more recently in \cite{troger2026}.} and a memory $M$, modelled as a qubit.
We will keep the notation general; however, when applying the results of this computation to the Wigner's friend experiment as studied above, $S$ refers to the spin-1/2 system measured by the friend Alice, $M$ denotes her memory, i.e.,~$A_I$, and $R$ refers to the orientation $A_R$ of Alice's lab. This von Neumann measurement represents Bob's description of Alice's measurement, where the orientation variables are described with respect to his frame.

Denote by $\vec J_S$ and $\vec J_R$ the angular momentum operators on $S$ and $R$ expressed relative to the external frame. Let us take the initial state to be
\begin{align}
    \ket{\Psi_0}= \left(\alpha\ket\up+\beta\ket\down\right)_S \otimes \ket 0_M \otimes \ket j_R ,
\end{align}
where $\ket\up_S$ and $\ket\down_S$ are eigenstates of $J_S^{(z)}$, the $z$-component of the angular momentum of $S$, and $\ket j_R$ is the maximal eigenstate of $J_R^{(z)}$, the angular momentum of the frame.  
Let us model the measurement via the interaction Hamiltonian
\begin{align} 
H = g\, H_{SR} \otimes Y_M ,
\ \text{with }
H_{SR} = -\frac{1}{2j{+}1}\bigl(2\,\vec J_S \cdot \vec J_R - j\bigr), \label{eq:H_SR}
\end{align}
where $Y_M = i(\lvert1\rangle\!\langle0| - \lvert0\rangle\!\langle1|)_M$ is the Pauli $\sigma_y$ operator on the memory. This interaction is rotationally invariant and therefore does not depend on the external reference frame.
After an interaction time $t = \pi/2g,$ the state is found to be
\begin{equation} \label{eq:psi_f}
\ket{\Psi_f}
= \left(\alpha\ket\up_S\ket0_M+\frac{2j}{2j+1}\beta\ket\down_S\ket1_M\right)\otimes \ket j_R+
\frac{\sqrt{4j{+}1}}{2j{+}1}\beta\lvert\Psi_{\mathrm{deg}}\rangle.
\end{equation}
For a large frame ($j\gg1$), the first term is arbitrarily close to the ideal measurement result $(\alpha\ket\up_S\ket0_M+\beta\ket\down_S\ket1_M)\otimes\ket j_R$ 
in which the reference frame remains unentangled, while the remaining term, where
\begin{equation}
\lvert\Psi_{\mathrm{deg}}\rangle
=
\frac{1}{\sqrt{4j{+}1}}\left(\ket\down_S\ket 0_M \ket j_R + \sqrt{2j}\ket\up_S\ket0_M\ket{j-1}_R-\sqrt{2j}\ket\up_S\ket1_M\ket{j-1}_R \right),
\end{equation}
is normalised,
encodes the inaccuracy of the measurement and the degradation of the frame through entanglement with the spin and the memory. The detailed computation can be found in Appendix \ref{app:computation}.

The measurement interaction inevitably induces some amount of entanglement between the spin system and the reference frame. This means that a measurement of the frame to check alignment to an external frame would lead to some decoherence of the memory and the spin. However, the entanglement can be made arbitrarily small by choosing a large enough $j$. Thus, any resulting decoherence associated with communicating or verifying the frame alignment can be made arbitrarily small by choosing a sufficiently large reference frame. Hence, there is no fundamental obstruction to maintaining alignment between isolated reference frames during a spin measurement and to performing such a measurement without decohering the superposition between the spin and the memory.

\section{A stronger proposal: no superposition of symmetry-invariant quantities}
\label{sec:Adash}

As mentioned in Section~\ref{sec:AMQM}, Adlam considers modification \textbf{A} a consequence of another, stronger, modification:
\begin{itemize}\setlength\itemsep{0em}
	\item \textbf{A'. Symmetry-invariant degrees of freedom are always definite:} In AMQM, symmetry-invariant degrees of freedom of \enquote{well-integrated} systems never go in superposition, thus they do not obey quantum theory.\footnote{The modification, as stated in the paper, reads: \enquote{for any well-integrated subsystem $U$ such that all of its parts currently share a reference frame, any property 
	$P$ of $U$ which is invariant under all relevant symmetry transformations will always have a well-defined value} \cite[p.\,12]{adlam2025wigners}.}
\end{itemize}
Adlam introduces modification \textbf{A'} as a speculative research direction, motivated by broader considerations concerning the relation between symmetry and quantum indefiniteness. Importantly, modification \textbf{A'} goes beyond explaining why observers in an EWFS experience definite outcomes and is offered instead as a general principle of a modified quantum theory that constrains which degrees of freedom can ever undergo superposition, in any physical setting. The principle, however, is not fully developed in Adlam's paper. In particular, the definition of a ``well-integrated'' system is not entirely specified.\footnote{Note that removing the ``well-integrated'' clause (``In AMQM, symmetry-invariant degrees of freedom never go in superposition, thus they do not obey quantum theory.'')  would simply put \textbf{A'} in direct contradiction with modification \textbf{B}. Modification \textbf{B} requires that, when Friend measures a spin in her isolated laboratory, the orientation of her lab relative to Wigner becomes entangled with the spin and, in particular, that this relative orientation enters a coherent superposition. But the orientation of Friend's lab relative to Wigner's is, by construction, invariant under a global rotation of both observers: it is a symmetry-invariant degree of freedom. In other words, the very degree of freedom that \textbf{B} requires to enter a superposition is exactly the kind \textbf{A'} requires to be definite.} 
(Clearly, since \textbf{A'} is supposed to ground \textbf{A}, observers need to be ``well-integrated''.) Adlam herself mentions two possibilities.

 The first is that the parts of a well-integrated system constantly interact and undergo \enquote{decoherence relative to one another}, such that \enquote{they all belong to the same reference frame} \cite[p.\,12]{adlam2025wigners}. It is unclear what it means for parts to undergo \enquote{decoherence relative to one another} and no mechanism is presented. Perhaps what is meant is that the information about this variable is leaked into the environment. However, it is well known that which degrees of freedom constitute ``the environment'' is not an absolute notion but is itself dependent on the real or presumed experimental capabilities of an observer. In Wigner's Friend, for example, the positions of the pointer variables of Friend's measurement instruments decohere with respect to Friend but are, \textit{by hypothesis}, in a coherent superposition with respect to Wigner, who can manipulate the entirety of Friend's lab. With this in mind, it seems that what is or is not ``well-integrated'' would itself be a relative notion and could not fulfil Adlam's ambition of finding an objective way to decide which variables have well-defined values.

 The second possibility that Adlam considers is that it is sufficient for the parts of a well-integrated system  to be \enquote{in continuous interaction} and that in fact \textit{all} interactions \enquote{always happen in a single definite way} \cite[p.\,14]{adlam2025wigners}. In the following, we take this as the definition of ``well-integrated'' and consider what \textbf{A'} would then imply.

Consider the simplest example of such a well-integrated system in a theory with translational invariance, in which relative distances are symmetry-invariant quantities. According to modification \textbf{A'}, such distances could never be in superposition. This would be astonishingly difficult to reconcile with well-established physical phenomena, as the description of atoms and molecules (whose parts are in constant interaction) relies critically on the delocalisation of electrons relative to nuclei, and hence on superpositions of relative positions. The hydrogen ground state---and indeed every stationary state of every atom---is precisely a coherent superposition of electron-nucleus relative positions. Chemistry, as it is currently understood, depends crucially on such superpositions, and the entire structure of the periodic table rests on them.

Adlam proposes two ways to overcome this difficulty~\cite{adlam2025wigners,adlam_email}. The first one is to claim that the relative distance between electrons and nuclei is not really an invariant but transforms under an additional symmetry, the second is to say that the observations that led us to believe in chemistry and nuclear and atomic physics are subject to an extension of modification \textbf{C}. Either move requires additional assumptions beyond those explicitly defended in Adlam’s proposal. The first amounts to postulating so-far unknown---hidden, that is---physical symmetries. While this may be a logical possibility, it seems unmotivated and, given the central role symmetries have played in the development of physical theories, we consider this rather implausible.

The second option, appealing to an analogue of modification \textbf{C}, amounts to denying that everyday chemical observations probe what they appear to probe, replacing them with unspecified hidden degrees of freedom. According to this view, just as Bob learns about the orientation of the spin in \textit{his} lab when he should just ask Alice what outcome she observed, experimenters measuring the statistics of electron positions around the nucleus probe \textit{some other} degree of freedom. Note that in this context the substitution cannot be a choice of the experimenters, as it is in the EWFS protocol: chemists do not deliberately record substitute variables. Here \textbf{C} must operate as a modification of the theory itself, a modification that introduces a radical disconnect between what experimenters think they measure and what they actually measure.

\section{Discussion}
\label{sec:discussion}

The no-go theorem of Bong \textit{et al.}~\cite{bong2020strong} establishes that the universal applicability of quantum theory, absoluteness of observed events (AOE), locality, and no-superdeterminism cannot all hold simultaneously. In attempts to preserve AOE,  people have considered two main options: (i) claim that quantum theory has limited applicability and the EWFS experiments cannot be performed, not even in principle; or (ii) keep the quantum mechanical predictions and accept the failure of one or more of the no-go theorem's other assumptions, via non-local, superdeterministic, or retrocausal effects. Adlam seems to propose a new way: draw on ideas from quantum reference frames to retain AOE, locality, no-superdeterminism, \emph{all the while} reproducing the statistics of quantum theory for an EWF scenario. However, as our analysis shows, her proposal does not in fact realise this combination.

\medskip

Adlam introduces three separate modifications to quantum theory and the protocol of an EWF experiment. Modification \textbf{A} explicitly enforces AOE, while the other modifications \textbf{B} and \textbf{C} enable the modified theory to reproduce, in specific contexts, the statistics of quantum theory. In this work, we have systematically analysed these modifications and their consequences and critiqued Adlam's motivations for introducing them.

We found modification \textbf{C} the most troubling. Essentially Adlam is proposing that the superobservers perform a different experiment than the one featured in the Bong \textit{et al.} no-go theorem. While in the standard EWF experiment, in some rounds the superobservers copy the friends' experimental outcomes, in Adlam's modified experiment the superobservers are only concerned with orientation degrees of freedom; see Table~\ref{tab:bong-vs-adlam}. She motivates this by claiming that there is an ambiguity about the variables featured in the no-go theorem, but there is no such ambiguity, as discussed in Section~\ref{sec:variables_in_EWFS}. In fact, the no-go theorem is not about orientations: it is completely agnostic about the nature of the systems measured by the friends. Adlam’s proposed resolution amounts to considering, within her theory, a different experiment that reproduces the statistics predicted by quantum theory for the EWF scenario, and interpreting the no-go theorem as applying to that experiment. 

Modification \textbf{B} is motivated by the claim that the isolation necessary to perform the EWFS requires the friends to be unable to detect any changes in orientation relative to the superobservers' frames, as comparing their orientation would spoil the superposition. As we discussed in Section~\ref{sec:isolation_shared_reference}, this is not the case. It is well-known that Friend can remain in a superposition relative to Wigner while the two communicate, as long as no information about the result of Friend's measurement leaves the sealed laboratory. We present a measurement model demonstrating how a totally isolated friend can measure the orientation of the spin relative to a physical reference frame in their lab, with controllable disturbance on this frame. This demonstrates that there is no fundamental limit to performing the EWFS on spins while keeping the frames aligned. The flipping of the observers postulated by modification \textbf{B} is thus not a natural consequence of isolation, but an independent assumption in conflict with standard quantum theory.

\medskip

It is worth noting a further, more fundamental limitation of Adlam's proposal: it is formulated entirely for measurements of spin orientation. Admittedly, one can imagine analogous protocols for other symmetry-variant quantities with modifications \textbf{B} and \textbf{C}: in a position-based EWFS, say, modification \textbf{B} would presumably entangle the location of the friends' laboratories with the measured particle's position and modification \textbf{C} would then require the superobservers to perform experiments on the positions of the systems and the friends. But nothing in the no-go theorem requires the friend to measure a symmetry-variant quantity. If they were to measure a symmetry-\textit{invariant} degree of freedom, such as the occupation number of an oscillator or whether an atom has decayed, no symmetry-variant quantity would be available to play the role that orientation plays in Adlam's construction. But quantum theory applies to such scenarios exactly as it does to spins; how AMQM handles them remains unspecified.

\medskip

The second half of Adlam’s paper explores a broader speculative programme linking the relational structure of quantum reference frames to relational interpretations of quantum mechanics. While this is an original direction to look for a principled way to impose limits on quantum theory, it seems to us that it cannot be taken particularly far. The proposal (which we call modification \textbf{A'}) is that symmetry-invariant quantities of ``well-integrated'' systems never undergo superposition. Adlam mentions two ways of understanding \enquote{well-integrated}, and we found either option to be lacking. If the parts of a system are well-integrated because they undergo ``decoherence relative to one another'', the notion inherits the observer-dependence of decoherence, and \textbf{A'} cannot objectively single out which variables have definite values. If they are well-integrated because they are in continuous interaction, then \textbf{A'} contradicts atomic and molecular physics: every stationary state of an atom is a coherent superposition of relative positions of parts in continuous interaction. To maintain \textbf{A'}, one must either postulate hidden symmetries, making the principle unfalsifiable, or extend modification \textbf{C} to everyday chemistry~\cite{adlam_email}.
\medskip

More generally, the proposal makes several claims about quantum theory that do not survive closer inspection. Contrary to one of her premises, quantum mechanics \textit{does} make predictions for variables defined in different frames, even if these frames are not in a definite relation.  In such cases, as we saw in Section~\ref{sec:nofact}, a well-defined probability distribution can still be obtained by treating reference frames quantum mechanically and, where appropriate, averaging over the relevant symmetry group \cite{Bartlett_2007,Giacomini_2017_covariance,Vanrietvelde_2018a,Hoehn_2019_trinity,delaHamette_2020}. Likewise, the claim that EWFS require such strong isolation that agents could not even notice changes in their own frame orientation is not supported by standard measurement theory and is not a requirement of the original formulations of these scenarios. 

While a cursory reading of Adlam’s paper may give the impression that quantum reference frames offer a resolution of the no-go theorems associated with extended Wigner’s Friend scenarios, our analysis shows that this is not the case: the resolution proposed in~\cite{adlam2025wigners} is not obtained from applying the quantum reference frames formalism within quantum theory. Instead, the resolution amounts to positing a modified theory in which a \textit{different} experiment has non-local statistics. Quantum reference frames, properly understood, provide systematic tools for handling agents without shared classical frames; they do not, however, solve the measurement problem~\cite{dibiagio2025talk,delahamette2026preparation}.

\begin{acknowledgments}
We thank Emily Adlam for useful clarifications about her proposal and for valuable comments on an earlier version of this comment. We further thank Timotheus Riedel for helpful comments on an earlier draft as well as Eric Cavalcanti for interesting discussions. Finally, we thank the QIT group at ETH for valuable feedback.

ACdlH acknowledges support from the Institute for Theoretical Studies at ETH Zurich through a Junior Research Fellowship.
This project was funded within the QuantERA II Programme that has received funding from the European Union’s
Horizon 2020 research and innovation programme under Grant Agreement No 101017733, and from the Austrian
Science Fund (FWF), projects I-6004 and ESP2889224 as well as Grant No.~I 5384.
\end{acknowledgments}

\bibliography{refs}

\pagebreak

\appendix

\section{Apparent violation of LF inequalities in AMQM}
\label{app:violation}

The correlation polytope defined by the LF inequalities is in general larger than the one defined by the Bell inequalities. However for two dichotomic measurements these two polytopes coincide~\cite{bong2020strong}. Here, violating a CHSH inequality means violating an LF inequality.

Let us rewrite the four states in~\eqref{eafs-states-new} as
\begin{equation}
  \ket{\Psi_{ac}}=\ket{\psi_{ac}}_{S_1A_RS_2C_R}\otimes \ket{a}_{A_I}\ket c_{C_I},
\end{equation}
for $a,c\in\{{+}1,{-}1\}$. Let us also introduce the basis
\begin{equation}
  \{|0\rangle\equiv \ket{\up\up},|1\rangle\equiv \ket{\up\down},|2\rangle\equiv \ket{\down\up},|3\rangle\equiv \ket{\down\down}\}
\end{equation}
for each of the pairs $S_1A_R$ and $S_2C_R$ so that we may write
\begin{equation}
 \begin{aligned}
\!\!\ket{\psi_{++}} = \frac{1}{\sqrt{2}}\bigl(\ket{00}+\ket{33}\bigr)\,,~
\ket{\psi_{+-}} = \frac{1}{\sqrt{2}}\bigl(\ket{01}+\ket{32}\bigr)\,,~
\ket{\psi_{-+}} = \frac{1}{\sqrt{2}}\bigl(\ket{10}+\ket{23}\bigr)\,,~
\ket{\psi_{--}} = \frac{1}{\sqrt{2}}\bigl(\ket{11}+\ket{22}\bigr)\,.~
\end{aligned}
\end{equation}
On each round, the variables $A_I$ and $C_I$ take on some definite value $a$ and $c$, respectively, and the friends and the spins are in one of the states $\ket{\Psi_{ac}}$. However, the superobservers do not know what the friends observed. So from their perspectives, Adlam’s evolution rule yields the separable mixture\footnote{Again, Adlam does not specify what the relevant frequencies of the friends' observations should be. Here we assume they follow the Born rule.}
\begin{align}
\rho
=
\frac{1}{4}\sum_{a,c\in\{{+}1,{-}1\}}
\ket{\Psi_{ac}}\!\bra{\Psi_{ac}}=\frac{1}{4}\sum_{ac\in\{{+}1,{-}1\}}
\ket{\psi_{ac}}\!\bra{\psi_{ac}}\otimes\ketbra{ac}{ac}.
\end{align}
One might expect that one cannot achieve any Bell-type violation given this mixture. However, this is not the case here. Indeed, let us define the following dichotomic observables on $S_1A_R$ and $S_2C_R$ in the basis $\{\ket i\}_{i\in\{0,\dots,3\}}$ given above:
\begin{align}
\hat Z:=\mathrm{diag}(1,1,-1,-1),
\qquad
\hat X:=
\begin{pmatrix}
0&0&0&1\\
0&0&1&0\\
0&1&0&0\\
1&0&0&0
\end{pmatrix},
\end{align}
and, with it, define the CHSH-like measurements
\begin{align}
    B_1:=\hat Z,\quad B_2:=\hat X,
\qquad
D_1:=\frac{\hat Z+\hat X}{\sqrt{2}},\quad D_2:=\frac{\hat Z-\hat X}{\sqrt{2}}.
\end{align}
Finally, define the CHSH operator
\begin{align}
M=\mathbb{1}_{A_IC_I}\otimes(B_1\otimes D_1 + B_1\otimes D_2 + B_2\otimes D_1 - B_2\otimes D_2)_{S_1A_RS_2C_R}.
\end{align}
One finds
\begin{equation}
\mathrm{tr}\,M\rho=2\sqrt{2}.
\end{equation}

Crucially, the observables $\hat Z$ and $\hat X$ act exclusively on the external degrees of freedom $S_1A_R$ and $S_2C_R$; they do \emph{not} probe the internal memory registers $A_I$ and $C_I$. In particular, when Bob receives input $x=1$, he measures the observable $B_1=\hat Z$ on the composite system $S_1A_R$. This measurement corresponds to measuring the spin of $S_1$ relative to \emph{Bob’s} frame and thus does \emph{not} probe Alice’s internal memory state, as implied by modification \textbf{C}. When Bob receives input $x=2$, he measures the observable $B_2=\hat X$, which probes coherence between different relative orientations of Alice’s laboratory. A similar interpretation applies to Divya’s measurements on Chidi’s laboratory.

Importantly, in neither setting do the superobservers' measurements  regard the friends’ internal configurations or memory registers encoding $A_I$ or $C_I$. This stands in contrast to the standard EWFS, where one of the superobservers’ measurement choices is explicitly taken to correspond to a direct readout of (or question about) what their friend has observed.

\section{Von Neumann spin measurement} \label{app:computation}

In this appendix, we give the explicit derivation of the post-measurement state $\ket{\Psi_f}$ given in Eq.~\eqref{eq:psi_f}, for the joint evolution of the spin-1/2 system $S$, the spin-$j$ reference frame $R$, and the memory qubit $M$ under the rotationally-invariant Hamiltonian of Eq.~\eqref{eq:H_SR}, using $\mathrm{SU(2)}$ recoupling theory.

We begin by rewriting $(\alpha\ket\up+\beta\ket\down)\lvert j\rangle$ in terms of eigenstates of the total angular momentum $J^2 = (\vec J_S + \vec J_R)^2$,
using Clebsch-Gordan coefficients:
\begin{align}
\lvert\uparrow\rangle\lvert j\rangle &= \lvert j+\tfrac12;\, j+\tfrac12\rangle, \\
\lvert\downarrow\rangle\lvert j\rangle
&=
\frac{1}{\sqrt{2j{+}1}}\lvert j+\tfrac12;\, j-\tfrac12\rangle
-
\sqrt{\frac{2j}{2j{+}1}}\lvert j-\tfrac12;\, j-\tfrac12\rangle.
\end{align}
Thus,
\begin{equation}
(\alpha\ket\up+\beta\ket\down)\lvert j\rangle
=
\alpha\lvert j+\tfrac12;\, j+\tfrac12\rangle
+
\frac{\beta}{\sqrt{2j{+}1}}\lvert j+\tfrac12;\, j-\tfrac12\rangle
-
\beta\sqrt{\frac{2j}{2j{+}1}}\lvert j-\tfrac12;\, j-\tfrac12\rangle.
\end{equation}
The Hamiltonian can be rewritten in terms of $J^2$ as
\begin{equation}
H = g\, \Pi_- \otimes \sigma_y ,
\end{equation}
where $\Pi_-$ is the projector onto the $J^2 = (j-\tfrac12)(j+\tfrac12)$ eigenspace. 
Indeed, note that
\begin{equation}
	J^2 = (\vec J_S + \vec J_R)^2 = J_S^2 + J_R^2 + 2\,\vec J_S \cdot \vec J_R
= \tfrac{3}{4} + j(j{+}1) + 2\,\vec J_S \cdot \vec J_R,
\end{equation}
so that
\begin{equation}
	2\,\vec J_S \cdot \vec J_R - j
= J^2 - \bigl(j+\tfrac12\bigr)\bigl(j+\tfrac32\bigr).
\end{equation}
It follows that $H_{SR}\lvert j+\tfrac12;\, m\rangle = 0$ and 
$H_{SR}\lvert j-\tfrac12;\, m\rangle = \lvert j-\tfrac12;\, m\rangle$.
The resulting time evolution is
\begin{equation}
U(t) = e^{-iHt}
=
(\mathbb 1 - \Pi_-)\otimes \mathbb 1_M
+
\Pi_- \otimes e^{-igY_M t}.
\end{equation}
Applying this to the initial state yields
\begin{equation}
U(t)\lvert\Psi_0\rangle
=
\Bigl(\alpha\lvert j+\tfrac12;\, j+\tfrac12\rangle
+
\frac{\beta}{\sqrt{2j{+}1}}\lvert j+\tfrac12;\, j-\tfrac12\rangle\Bigr)\lvert0\rangle_M
-
\beta\sqrt{\frac{2j}{2j{+}1}}\lvert j-\tfrac12;\, j-\tfrac12\rangle e^{-igY_M t}\lvert0\rangle_M.
\end{equation}
For an interaction time $t = \pi/2g$, this becomes
\begin{equation}
\lvert\Psi_f\rangle
=
\Bigl(\alpha\lvert j+\tfrac12;\, j+\tfrac12\rangle
+
\frac{\beta}{\sqrt{2j{+}1}}\lvert j+\tfrac12;\, j-\tfrac12\rangle\Bigr)\lvert0\rangle_M
-
\beta\sqrt{\frac{2j}{2j{+}1}}\lvert j-\tfrac12;\, j-\tfrac12\rangle\lvert1\rangle_M.
\end{equation}

Inverting the Clebsch--Gordan decomposition,
\begin{align}
\lvert j+\tfrac12;\, j-\tfrac12\rangle
&=
\frac{1}{\sqrt{2j{+}1}}\lvert\downarrow\rangle\lvert j\rangle
+
\sqrt{\frac{2j}{2j{+}1}}\lvert\uparrow\rangle\lvert j-1\rangle, \\
\lvert j-\tfrac12;\, j-\tfrac12\rangle
&=
-\sqrt{\frac{2j}{2j{+}1}}\lvert\downarrow\rangle\lvert j\rangle
+
\frac{1}{\sqrt{2j{+}1}}\lvert\uparrow\rangle\lvert j-1\rangle,
\end{align}
we obtain
\begin{equation}
\lvert\Psi_f\rangle=
\Bigl[\alpha\lvert\uparrow\rangle\lvert j\rangle
+
\frac{\beta}{2j{+}1}\bigl(\lvert\downarrow\rangle\lvert j\rangle
+
\sqrt{2j}\,\lvert\uparrow\rangle\lvert j-1\rangle\bigr)\Bigr]\lvert0\rangle_M 
+
\frac{2j}{2j{+}1}\beta
\bigl[\lvert\downarrow\rangle\lvert j\rangle
-
\frac1{\sqrt{2j}}\lvert\uparrow\rangle\lvert j-1\rangle\bigr]\lvert1\rangle_M .
\end{equation}
This state corresponds to Eq.~\eqref{eq:psi_f}.

\section{Correlation between observables in different quantum reference frames} \label{app:QRFs}

In this appendix, we show how to extract predictions about observables defined with respect to different frames using the perspectival and/or perspective-neutral framework. This allows us to extract information when the orientation of one frame is in some known quantum state (possibly entangled with other systems) with respect to another one.

Despite the somewhat heavy notation, the argument rests on a simple observation, namely, that ``the observable $O$ of system $S$ in $A$'s frame'' corresponds to a unique (symmetry-invariant) relational observable. Concretely, for the translation group, ``the position of particle 1 in the frame in which $A$ is at the origin'' is simply the displacement $\hat{x}_1-\hat{x}_A$ in all other frames.

If the orientation of one frame relative to another is in a known quantum state, possibly entangled with other systems, this information is encoded in a perspectival state $\ket{\psi} \in \H_{|A}$ describing the remaining systems from $A$'s frame. Expectation values of any observables defined relative to $A$'s frame are computed directly on $\ket{\psi}$ in the usual way, and correlations between observables defined relative to different frames are obtained using the frame-change map introduced below. Equivalently, the same predictions can be computed from the perspective-neutral state $\kket{\psi}\in\H^\phys$, obtained by group averaging, on which any relational observable can be evaluated. The equivalence between the two pictures is explained below.\\

Let us consider two systems $S_1$ and $S_2$ and two ideal frames $A$ and $B$ \cite{delahamette2021perspectiveneutralapproachquantumframe}, all transforming according to some group $G$. We have the kinematical Hilbert space $\H^\kin=\H_1\otimes\H_2\otimes\H_A\otimes\H_B$ and the projector on the physical Hilbert space
\begin{equation}
\Pi^\phys=\frac1{|G|}\sum_{g\in G}U(g),
\end{equation}
where $U(g)=U_1(g)\otimes U_2(g)\otimes U_A(g)\otimes U_B(g),$ with $U_A$ and $U_B$ regular representations and $U_1$ and $U_2$ some unitary representations of $G$.

We can express any perspective-neutral state $\kket\psi\in\H^\phys$ relative to frame $A$ by conditioning on $A$ being at the \enquote{origin}, that is, by conditioning on the state corresponding to the unit element $e$ of $G$. Namely, there is an isomorphism \begin{equation} \label{eq:reduction map}
\begin{aligned}
\mathcal R_{|A}:
\H^\phys&\longto \H_{|A}\\
\kket{\psi}&\longmapsto\sqrt{\smash{|}G\smash{|}} \;{}_{A}\langle{e}\kket\psi,
\end{aligned}
\end{equation}
called the \emph{reduction map}.\footnote{Note that Eq.\,\eqref{eq:reduction map} holds for compact groups. For non-compact groups one introduces a physical inner product to replace the kinematical one and ensure normalisation of states on the physical and relational Hilbert spaces (see \cite{delahamette2021perspectiveneutralapproachquantumframe} for more details).} The Hilbert space $\H_{|A}$ describes the other systems in the frame aligned to $A$. The inverse map,
\begin{equation}
\begin{aligned}
\mathcal R_{|A}^{-1}:
\H_{|A}&\longto\H^\phys\\
\ket{\psi}&\longmapsto \sqrt{\smash{|}G\smash{|}}\cdot \Pi^\phys\left(\ket{e}_A\otimes\ket{\psi}\right),
\end{aligned}
\end{equation}
is called the \textit{relativisation map}.

We can use the reduction and relativisation maps to express \textit{variables defined in a frame} as frame-independent, \textit{invariant} variables.
Let $O_{1|A}\in\mathcal L(\H_{|A})$ be \textit{some} observable on particle 1 expressed in $A$'s frame. Then this actually corresponds to the frame-independent relational observable $O_{1A}\in\mathcal L(\H^\phys)$, where
\begin{equation}
O_{1A}:=\mathcal R^{-1}_{|A}O_{1|A}\mathcal R_{|A}.
\end{equation}
For any state $\ket\psi_{|A}\in\H_{|A}$ these observables satisfy
\begin{equation}
\matrixel{\psi}{O_{1|A}}{\psi}_{|A}=\bbra\psi O_{1A}\kket\psi.
\end{equation}
Conversely, one can take \textit{any} frame-independent observable $O\in\mathcal L(\H^\phys)$ and see how it would look in a frame by using the inverse map.

After these basic considerations, one can quickly obtain correlations between observables defined in different frames. In particular, let $O_{1|A}$ and $O_{2|B}$ be observables of $S_1$ defined in $A$'s frame and of $S_2$ defined in $B$'s frame, respectively, and let $O_{1A},O_{2B}\in\mathcal L(\H^\phys)$ be the corresponding relational observables. Then the correlations are simply given by
\begin{equation}
C_{\psi}(O_{1|A}O_{2|B})\equiv\bbra\psi O_{1A}O_{2B}\kket\psi-\bbra\psi O_{1A}\kket\psi\bbra\psi O_{2B}\kket\psi
\end{equation}
in the usual way.

Alternatively, and equivalently, we can use frame change maps to translate an observable expressed relative to frame $A$ to one in frame $B$ via the frame change map
$V^{A\to B} = \mathcal R_{|B}\mathcal R_{|A}^{-1}: \H_{|A} \to \H_{|B}.$
This immediately leads to a family of invariance relations equating expectation values in $\H_{|A}$ with expectation values in $\H_{|B}$. For example,
\begin{equation}
\bra{\psi} O_{|A}\ket\psi{}_{|A} = \bra{\psi}{O_{|A}}^{A\to B}\ket{\psi}{}_{|B}
\end{equation}
where $\ket{\psi}_{|B}=V^{A\to B}\ket\psi_{|A}$ and ${O_{|A}}^{A\to B}=V^{A\to B}O_{|A}V^{B\to A}$ are defined in $\H_{|B}$.

Moreover, one can check that 
\begin{equation}
C_{\psi}(O_{1|A}O_{2|B})=\bra\psi O_{1|A}O_{2|B}^{B\to A}\ket\psi_{|A}-\bra\psi O_{1|A}\ket\psi_{|A}\,  \bra{\psi} O_{2|B}^{B\to A}\ket{\psi}_{|A}
\end{equation}
and, analogously,
\begin{equation}
C_{\psi}(O_{1|A}O_{2|B})=\bra\psi O_{1|A}^{A\to B}O_{2|B}\ket\psi_{|B}-\bra\psi O_{1|A}^{A\to B}\ket\psi_{|B}\,  \bra{\psi} O_{2|B}\ket{\psi}_{|B},
\end{equation}
allowing one explicitly to compute correlations between variables expressed in different frames.\\

Let us consider the concrete simple case where $G=\ZZ_2$, and $\H_1\cong\H_2\cong \H_A\cong\H_B\cong\CC^2$, each carrying the regular representation, with $\ket{e}=\ket{\up}$ and $\ket{a}=\ket{\down}$. The projector on the physical Hilbert space is \begin{equation}
\Pi^\phys=\frac12+\frac12X\otimes X\otimes X\otimes X,
\end{equation}
where $X=\ketbra{{\up}}{{\down}}+\ketbra{{\down}}{{\up}}$. Let $Z_{1|A}$ be the direction of $S_1$ in $A$'s frame. In the physical Hilbert space this becomes 
\begin{equation}
	Z_{1A}=Z_1Z_A=Z_1\otimes\mathbb{1}_2\otimes Z_A\otimes \mathbb 1_B.
\end{equation}

Similarly, $Z_{2|B}\in\mathcal L(\H_{|B})$ corresponds to $Z_{2B}=Z_2 Z_B \in\mathcal L(\H^\phys)$. What does the operator $Z_{2|B}\in\mathcal{L}(\H_{|B})$ for \enquote{the orientation of $S_2$ in Bob's frame} look like in Alice's frame?

We obtain this operator in Alice's frame by first mapping back to the corresponding physical observable and then reducing to Alice's perspective:
\begin{align}
    Z_{2|B}^{B\to A}=\mathcal{R}_{|A}(\mathcal{R}_{|B}^{-1}Z_{2|B}\mathcal{R}_{|B})\mathcal{R}_{|A}^{-1} = \mathcal{R}_{|A}Z_{2B}\mathcal{R}_{|A}^{-1}=\mathcal{R}_{|A}(Z_2Z_B)\mathcal{R}_{|A}^{-1}=\mathcal{R}_{|A}Z_2 \mathcal{R}_{|A}^{-1}\mathcal{R}_{|A}Z_B\mathcal{R}_{|A}^{-1} = Z_{2|A}Z_{B|A},
\end{align}
where we used $Z_{2B}=Z_2Z_B$ and $\mathcal{R}_{|A}^{-1}\mathcal{R}_{|A}=\mathbb{1}$.
This illustrates that we can freely transform variables defined relative to one frame to variables relative to another frame.
\end{document}